\documentclass[floatfix,twocolumn,showpacs,preprintnumbers,amsmath,amssymb]{revtex4}

\usepackage{graphicx}
\usepackage{dcolumn}
\usepackage{bm}

\begin{document}

\title{
Magnetic Properties of the Novel Low-Dimensional Cuprate Na$_{\rm
5}$RbCu$_{\rm 4}$(AsO$_{\rm4}$)$_{\rm 4}$Cl$_{\rm 2}$}

\author{J. A. Clayhold}
\affiliation{Department of Physics and Astronomy, Kinard Laboratory, Clemson
University,
Clemson, SC 29634}
\author{M. Ulutagay-Kartin and S.-J. Hwu}
\affiliation{Department of Chemistry, Clemson
University,
Clemson, SC 29634}
\author{H.-J. Koo and M.-H. Whangbo}
\affiliation{Department of Chemistry, North Carolina State University, Raleigh, NC
27695-8204}
\author{A. Voigt}
\affiliation{Center for Simulational Physics, Department of Physics and Astronomy,
University of Georgia, Athens, GA 30602}
\author{K. Eaiprasertsak}
\affiliation{Department of Materials Science,  Clemson University, Clemson, SC 29634}

\date{\today}
\begin{abstract}
The magnetic properties  of a new  compound, \mbox{Na$_{\rm
5}$RbCu$_{\rm 4}$(AsO$_{\rm 4}$)$_{\rm 4}$Cl$_{\rm 2}$}, are
reported. The material has a layered structure comprised of square
Cu$_4$O$_4$ tetramers.  The Cu ions are divalent and the system
behaves as a low-dimensional S=1/2 antiferromagnet.  Spin exchange in
Na$_{\rm 5}$RbCu$_{\rm 4}$(AsO$_{\rm 4}$)$_{\rm 4}$Cl$_{\rm 2}$ appears to be quasi-two-dimensional
and non-frustrated.  Measurements
of the bulk magnetic susceptibility and heat capacity  are consistent with low-dimensional magnetism.    The compound has an
interesting, low-entropy, magnetic transition at T = 17 K.
\end{abstract}

\pacs{75.30.Et, 75.40.Cx, 75.50.-y, 75.50.Ee}

\maketitle

\section*{Introduction}

The occurrence of high-temperature superconductivity in doped spin-1/2 square planar
antiferromagnets has stimulated  the quest for new families of low-dimensional magnetic materials.  
We have studied and report here the first results from a new compound,  Na$_{\rm 5}$RbCu$_{\rm 4}$(AsO$_{\rm
4}$)$_{\rm 4}$Cl$_{\rm 2}$, whose spin exchange interactions are confined to
two-dimensional layers.

\begin{figure}
\begin{center}
\includegraphics[width=2.4in]{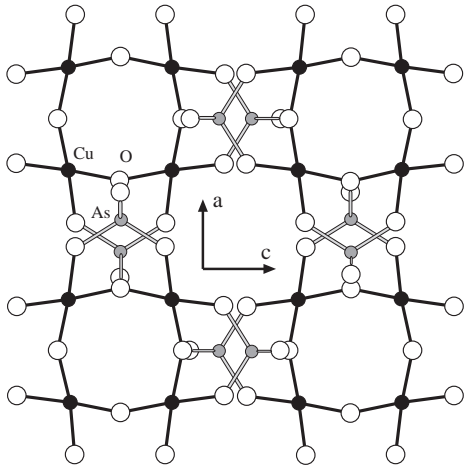}
\end{center}
\protect\caption{
Perspective view of a Cu$_{\rm
4}$(AsO$_{\rm 4}$)$_{\rm 4}$ layer of Na$_{\rm 5}$RbCu$_{\rm
4}$(AsO$_{\rm 4}$)$_{\rm 4}$Cl$_{\rm 2}$. The layer is contained
in the ac-plane.
\label{planes}}
\end{figure}
\mbox{Na$_{\rm 5}$RbCu$_{\rm 4}$(AsO$_{\rm 4}$)$_{\rm 4}$Cl$_{\rm
2}$} is a nearly-tetragonal insulating magnetic material with a
remarkable crystal structure\cite{NaRb_JACS}.  It is a
two-dimensional, layered compound with a square-planar arrangement
of copper and oxygen ions. The copper valence state is 2+,
so the Cu ions are magnetic with spin 1/2.  The copper and
oxygen ions form square tetramer units, Cu$_{\rm 4}$O$_{\rm
4}$, which are connected by the AsO$_4$ bridging units to form
nearly-tetragonal two-dimensional layers, as illustrated in
Fig.~\ref{planes}. Thus the spin exchange of Na$_{\rm5}$RbCu$_{\rm
4}$(AsO$_{\rm 4}$)$_{\rm 4}$Cl$_{\rm 2}$ is expected to be
essentially two-dimensional in nature (see below for details).

\section*{Experiment}

Single-crystals of \mbox{Na$_{\rm 5}$RbCu$_{\rm 4}$(AsO$_{\rm
4}$)$_{\rm 4}$Cl$_{\rm 2}$} were synthesized by conventional solid
state reaction using molten-salt methods.  Details of the
synthesis are reported elsewhere\cite{NaRb_JACS}.  The crystals
are transparent, blue and plate-like.  Typical crystal dimensions
are 1~mm $\times$ 1~mm $\times$ 200~$\mu$m.  The crystal structure
was determined by employing single-crystal X-ray diffraction
techniques.

The materials were characterized by specific heat and magnetization
measurements.  Special care was required for these measurements because
of the small size of the sample materials.  We used a Quantum Design
SQUID magnetometer for the magnetization studies and were careful to
first measure an empty sample holder---with mounting grease---each time
before affixing the sample.  Even with the largest sample (2~mm $\times$
2.3~mm $\times$ 500~$\mu$m) of \mbox{Na$_{\rm 5}$RbCu$_{\rm
4}$(AsO$_{\rm 4}$)$_{\rm 4}$Cl$_{\rm 2}$}, the diamagnetic background
signal from the tiny amount of silicone vacuum grease was nearly 20 \%
of the measured signal, necessitating the careful subtraction.
Nevertheless, it was possible to determine the anisotropy of the
magnetization with good precision.

The specific heat measurements presented an even greater challenge.
The largest crystals weigh only 3 milligrams and have a total heat
capacity of only 25~$\mu$J/K at T~=15~K. Nevertheless, we have
constructed a calorimeter capable of resolving bulk heat capacities to
a resolution of nearly 100~nanoJoules-per-Kelvin.  With this apparatus,
it was possible to measure single-crystal samples.

\begin{figure}
\includegraphics[width=2.9in]{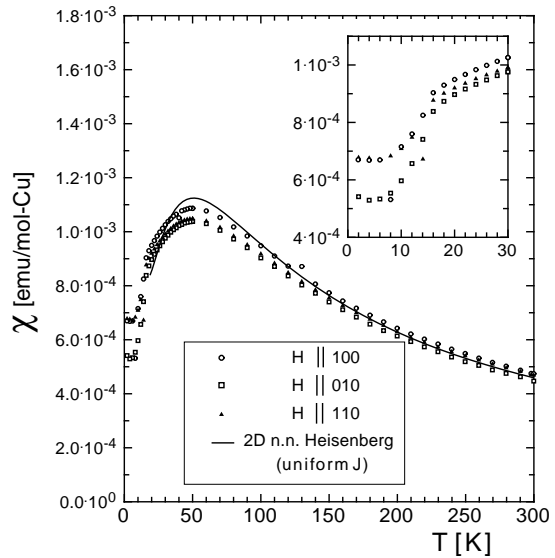}
\protect\caption{ The magnetic susceptibility of \mbox{Na$_{\rm
5}$RbCu$_{\rm 4}$(AsO$_{\rm 4}$)$_{\rm 4}$Cl$_{\rm 2}$} for
different orientations of the magnetic field with respect to the
crystal axes, as indicated.   The solid curve shows the temperature dependence
of the magnetic susceptibility of the two-dimensional Heisenberg model with $J$~=~4.77~meV.  The inset shows the
antiferromagnetic phase transition at T~=~17~K. \label{chiRbNa}}
\end{figure}

Susceptibility data for \mbox{Na$_{\rm 5}$RbCu$_{\rm 4}$(AsO$_{\rm
4}$)$_{\rm 4}$Cl$_{\rm 2}$} are shown in Fig.~\ref{chiRbNa}. The
measurements were carried out with the magnetic field applied
along different symmetry axes of the crystal.  The magnetic
susceptibility is slightly higher when the magnetic field is
applied along the a- or c-directions than along the b-direction.  No
hysteresis was detected at any temperature.  Above T~=~100 K, it
follows a Curie-Weiss law with an antiferromagnetic Curie-Weiss
temperature of -86~K with $g$~=~1.97, consistent with
crystal-field quenching of the orbital angular momentum.  At
T~=~50~K, the susceptibility reaches a broad maximum before
decreasing at lower temperatures.  The solid curve shows a
qualitative fit to a high-temperature series expansion\cite{highTser1,highTser2} for the magnetic susceptibility of the 
uniform two-dimensional nearest-neighbor Heisenberg model, 
yielding an effective  single value for the magnetic coupling, $J =$~4.77~meV.   

The susceptibility data shows a phase transition  at T~=~17~K.  As
shown in the inset to Fig.~\ref{chiRbNa}, the susceptibility drops
by approximately 30 to 40 \% at the transition temperature and
becomes much more anisotropic.  Below the transition, the
susceptibility is smallest when measured with the magnetic field
aligned perpendicular to the layers of the crystal.  It seems
likely that the magnetic transition upon cooling through 17~K
involves the onset of antiferromagnetic order.

\begin{figure}
\includegraphics[width=2.6in]{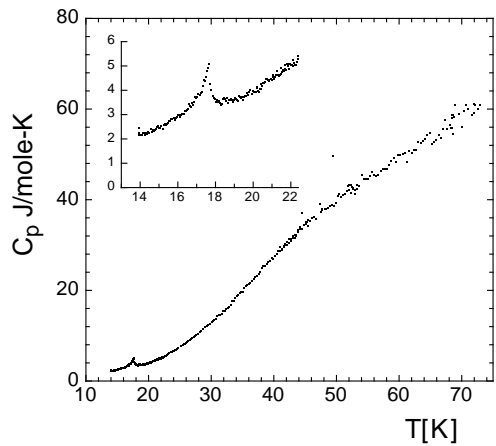}
\protect\caption{
Specific heat of \mbox{Na$_{\rm 5}$RbCu$_{\rm 4}$(AsO$_{\rm 4}$)$_{\rm
4}$Cl$_{\rm 2}$}.  The data are shown per mole of Cu$^{\rm 2+}$.  There
is a second-order phase transition at T~=~17~K, as shown more clearly
in the inset.  The total entropy involved in the
transition is  small, amounting to less than 10~\% of the
free-spin entropy-per-mole, $R\ln{2}$.
\label{CpRbNa}}
\end{figure}

Specific heat measurements taken on single crystals of
\mbox{Na$_{\rm 5}$RbCu$_{\rm 4}$(AsO$_{\rm 4}$)$_{\rm 4}$Cl$_{\rm
2}$} are shown in Fig.~\ref{CpRbNa}.  From these data, we find
a Debye temperature of 320~K. We also observed a second-order
phase transition at T~=~17~K as seen clearly in the inset to
Fig.~\ref{CpRbNa}. The transition temperature seen in the specific
heat corresponds to the temperature at which we observe the phase
transition in the magnetic susceptibility.  We also find that the
entropy loss per spin upon cooling through the transition amounts
to only a small fraction, less than 10 \%, of the total free spin
entropy, $R\ln{2}$, per Cu$^{\rm 2+}$.

 We performed a careful search for  a
phase transition, magnetic or otherwise, between 20~K and room
temperature using  ac calorimetry.  No transition was
found.  Additionally, the susceptibility was
measured at T~=~4.2~K in magnetic fields up to 33~T using a cantilever susceptometer 
to detect metamagnetic or spin-flop transitions at the
National High Magnetic Field Laboratory.   No field-induced transitions were found.

\section*{Discussion}

Because of its layered structure, the magnetic properties of
\mbox{Na$_{\rm 5}$RbCu$_{\rm 4}$(AsO$_{\rm 4}$)$_{\rm 4}$Cl$_{\rm
2}$} should be governed primarily by the spin exchange
interactions between neighboring Cu$^{2+}$ ions within the same
Cu$_{\rm 4}$(AsO$_{\rm 4}$)$_{\rm 4}$ layer. Spin exchange between
the layers of the compound should be very small because the
interlayer O$^{...}$O distances are long (greater than 6.1~{\AA}).
In addition, the stacking of the layers is staggered (see
Ref.~\onlinecite{NaRb_JACS} for details) in a way that frustrates
inter-layer exchange.

\begin{figure}
\includegraphics[width=1.8in]{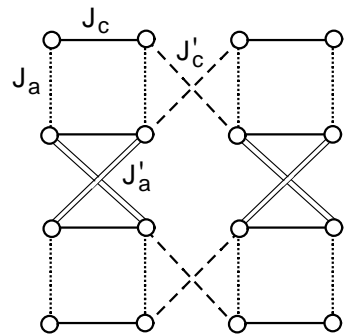}
\protect\caption{Non-frustrated Heisenberg antiferromagnetic spin exchange interactions
between neighboring Cu$^{2+}$ ions in the Cu$_{\rm 4}$(AsO$_{\rm
4}$)$_{\rm 4}$ layers of Na$_{\rm 5}$RbCu$_{\rm 4}$(AsO$_{\rm
4}$)$_{\rm 4}$Cl$_{\rm 2}$. J$_{\rm a}$ and J$_{\rm c}$ are
antiferromagnetic superexchange couplings within a single
tetramer; J$^\prime_{\rm a}$ and J$^\prime_{\rm c}$ are
antiferromagnetic super-superexchange couplings between adjacent
tetramers. 
\label{couplings}}
\end{figure}

Within a single Cu$_{\rm 4}$(AsO$_{\rm 4}$)$_{\rm 4}$ layer,
adjacent Cu$_{\rm 4}$O$_{\rm 4}$ tetramers are connected by
AsO$_{\rm 4}$ bridging units. It is, however, unlikely that
superexchange coupling via the As-O bonds is important because
even the P-O bonds in (VO)$_{\rm 2}$P$_{\rm 2}$O$_{\rm 7}$
are not an efficient medium for spin exchange\cite{KW_PO}
although the P 3s/3p orbitals overlap more
strongly with the O 2s/2p orbitals than do the As 4s/4p orbitals.
The spin exchange within each tetramer should take place via the
Cu-O-Cu superexchange paths. Thus there are two intra-tetramer
coupling constants, J$_{\rm a}$ and J$_{\rm c}$, to consider
because the a- and c-directions are not equivalent (see
Fig.~\ref{couplings}).

\begin{figure}
\begin{center}
\includegraphics[width=2.4in]{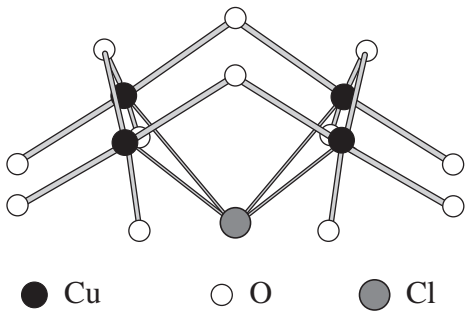}
\end{center}
\protect\caption{
Close-up view of a single Cu$_{\rm 4}$O$_{\rm 4}$ tetramer showing the bent Cu--O--Cu bonds.  
The bond angle is close to 110$^\circ$, causing the intratetramer superexchange couplings to be much smaller than in
cuprates having straight, 180$^\circ$ Cu--O--Cu bonds.  
\label{buckled}}
\end{figure}

As shown in Fig.~\ref{buckled}, the Cu$_{\rm 4}$O$_{\rm 4}$ tetramer units
are buckled.  The Cu--O--Cu bonds are far from straight, the bond angle being 110.9$^\circ$ along the a-axis
and 108.7$^\circ$ along the c-axis 
instead of 180$^\circ$.  This signficantly reduces the strength of the superexchange
coupling between adjacent copper ions, which can exceed 120~meV in compounds having colinear Cu--O--Cu bonds\cite{cuprateJ}.

Significant inter-tetramer interactions are
possible through the Cu-O$^{...}$O-Cu super-superexchange paths if
the Cu-O$^{...}$O-Cu bonds are close to linear and the O$^{...}$O
contact distances are short\cite{KW_SSE}. Analysis of the layer
structure suggests that such super-superexchange interactions
should occur between adjacent tetramers along the in-plane
diagonals. There are two inter-tetramer coupling constants,
J$^\prime_{\rm a}$ and J$^\prime_{\rm c}$, to consider (see
Fig.~\ref{couplings}).

Values for the four
different Heisenberg
couplings can be estimated on the basis of spin dimer
analysis\cite{KW_SSE,KW_PO}. This method calculates the hopping
integral, $t$, for the spin dimer\cite{variables} (i.e., the structural
unit containing two adjacent spin sites) corresponding to an exchange
path on the basis of tight-binding molecular orbital calculations.
For antiferromagnetic spin exchange, the coupling constant $J$ is
given by $J = \frac{4t^2}{U_{\rm eff}}$, where $U_{\rm eff}$ is
the effective onsite repulsion.  It was possible to estimate the value
of U$_{\rm eff} =$ 1.5 eV for our calculations involving Cu
$d_{x^2-y^2}$ orbitals  by comparing a set of experimental $J$
values with the corresponding calculated $t^2$ values for Nd$_{\rm
2}$CuO$_{\rm 4}$ and La$_{\rm 2}$CuO$_{\rm 4}$.\cite{perovskite}
Thus from the hopping integrals calculated for the four exchange paths
shown in Fig.~\ref{couplings}, their spin-exchange coupling
constants are estimated as follows: J$_{\rm a} \approx$ 6
meV, J$_{\rm c} \approx$ 3 meV, J$^\prime_{\rm a}\approx$ 6 meV,
and J$^\prime_{\rm c}\approx$ 5 meV.

Our calculations indicate that inter-tetramer interactions are as
important as the intra-tetramer interactions. We found  that the
spin exchange interactions through other spin exchange paths are
negligible. Thus the spin exchange interactions in Na$_{\rm
5}$RbCu$_{\rm 4}$(AsO$_{\rm 4}$)$_{\rm 4}$Cl$_{\rm 2}$ cannot be approximated by a model
of weakly interacting tetramers.   The calculations indicate that spin-exchange in 
Na$_{\rm 5}$RbCu$_{\rm 4}$(AsO$_{\rm 4}$)$_{\rm 4}$Cl$_{\rm 2}$ is quasi-two-dimensional
and non-frustrated.

As described above, ac calorimetry confirmed the absence of any
cooperative phase transition above 20~K. Only when the material
was cooled below 17~K were the spins observed to order
magnetically.  This phase transition at 17~K has some peculiar
features which appear to distinguish it from conventional
antiferromagnetic transitions. The first is that the observed
ordering temperature is much smaller, by a factor of five, than
the measured Curie-Weiss temperature, $|\theta|$~=~86~K.  This is
probably understandable because of the anisotropy of the
spin-exchange in Na$_{\rm 5}$RbCu$_{\rm 4}$(AsO$_{\rm 4}$)$_{\rm
4}$Cl$_{\rm 2}$:  the magnetic ordering necessarily involves the
much weaker magnetic coupling along the b-direction.

We note that the total entropy involved in the
second-order transition is  smaller---by an order of
magnitude---than is common for three-dimensional antiferromagnetic-ordering
transitions. It is obvious that the transition {\em does} involve
some sort of spin ordering because of its effects on the magnetic
susceptibility (Fig.~\ref{chiRbNa}, inset), which means that the
low entropy must be accounted for. The most likely explanation would
be that much of the spin degeneracy had already been lifted at
higher temperatures as the spin degrees of freedom develop
low-dimensional correlations upon cooling below $T \approx J/k_B$.

In summary, Na$_{\rm 5}$RbCu$_{\rm 4}$(AsO$_{\rm 4}$)$_{\rm
4}$Cl$_{\rm 2}$ is a new layered cuprate antiferromagnet, in which
spin exchange interactions are largely confined to 
well-separated layers.   The bilinear, Heisenberg spin exchange interactions appear to be
non-frustrated.

\begin{acknowledgments}
 The authors are are indebted to Drs.\ Stephen Nagler (Oak Ridge
National Laboratory), Malcolm Skove (Clemson University), and
Roman Movshovich (Los Alamos National Laboratory) for useful
conversations.  Financial support from National Science Foundation
 for the purchase of a SQUID magnetometer (CHE-9808044) and research
(DMR-0077321) is gratefully
acknowledged. In addition, the authors wish to thank Bruce Brandt
and Donavan Hall for valuable assistance at the National High
Magnetic Field Laboratory (NHMFL). The NHMFL is supported by the
NSF and the State of Florida.  The work at North Carolina State
University was supported by the Office of Basic Energy Sciences,
Division of Materials Sciences, U. S. Department of Energy, under
Grant DE-FG05-86ER45259.
\end{acknowledgments}
\bibliography{NaRb}

\begin{thebibliography}{8}
\bibitem[{var()}]{variables}
\bibinfo{note}{In refs. \onlinecite{KW_PO}, \onlinecite{KW_SSE}, and \onlinecite{perovskite}, the
variable, $\Delta e = 2t$, is used}.
\end{thebibliography}
\end{document}